\documentclass[reprint,amsmath,amssymb, aps, twocolumn
]
{revtex4}

\usepackage{graphicx}
\usepackage{dcolumn}
\usepackage{bm}

\begin{document}

\preprint{APS/123-QED}

\title{A Global Fit Determination of Effective  $\Delta m_{31}^2$ \\
from Baseline Dependence of Reactor $\bar{\nu}_e$ Disappearance}

\newcommand{\RCNS}{Research Center for Neutrino Science, Tohoku University, Sendai, 980-8578, Japan}
\newcommand{\TMU}{Department of Physics, Tokyo Metropolitan University, Tokyo, 192-0397, Japan}

\affiliation{\RCNS}
\affiliation{\TMU}

\author{T.J.C.~Bezerra}
\email{thiago@awa.tohoku.ac.jp}
\affiliation{\RCNS}

\author{H.~Furuta}
\email{furuta@awa.tohoku.ac.jp}
\affiliation{\RCNS}

\author{F.~Suekane}
\email{suekane@awa.tohoku.ac.jp}
\affiliation{\RCNS}

\author{T.~Matsubara}
\email{matsubara@hepmail.phys.se.tmu.ac.jp}
\affiliation{\TMU}

\date{\today}

\begin{abstract}
Recently, three reactor neutrino experiments, Daya Bay, Double Chooz and RENO have directly measured the neutrino mixing angle $\theta_{13}$.  
In this paper, another important oscillation parameter, effective $\Delta m_{31}^2$ 
(= $\Delta \tilde{m}_{31}^2$) is measured using baseline dependence of the reactor neutrino disappearances.    
A global fit is applied to publicly available data and 
$\Delta \tilde{m}_{31}^2 =  2.95^{+0.42}_{-0.61}  \times 10^{-3}$~eV$^2$,
~$\sin^22\theta_{13} = 0.099^{+0.016}_{-0.012}$ are obtained by setting both parameters free.
This result is complementary to  $\Delta \tilde{m}_{31}^2$ to be measured by spectrum shape analysis.
The measured $\Delta \tilde{m}_{31}^2$ is consistent with $\Delta \tilde{m}_{32}^2$ measured by $\nu_{\mu}$ disappearance in MINOS, T2K and atmospheric neutrino experiments within errors. 
The minimum $\chi^2$ is small, which means the results from the three reactor neutrino experiments are consistent with each other. 
\end{abstract}

\maketitle

\section{\label{sec:Introduction} Introduction}

Neutrino oscillation is a phenomenon which is not accounted for by the Standard Model of elementary particles, which assumes neutrinos as massless. 
There are six parameters in standard three flavor neutrino oscillation~\cite{PDG10}: three mixing angles between flavor eigenstates and mass eigenstates ($\theta_{12}$, $\theta_{13}$ and $\theta_{23}$), one CP violating imaginary phase ($\delta$), and two independent squared mass differences: $\Delta m^2_{jk} \equiv m^2_j-m^2_k$, where $m_i$ are neutrino masses ($m_1$, $m_2$, $m_3$) of the three mass eigenstates ($\nu_1$, $\nu_2$, $\nu_3$) which correspond to the largest component of ($\nu_e$, $\nu_{\mu}$, $\nu_{\tau}$), respectively. 
$\theta_{12}$ and $\Delta m_{21}^2$ have been measured by solar neutrino disappearance experiments 
$(\nu_e \to \nu_e)$ and long baseline reactor neutrino disappearance experiments
$(\bar{\nu}_e \to \bar{\nu}_e)$.
$\theta_{23}$ and $\Delta \tilde{m}^2_{32}$ have been measured by $(\nu_{\mu} \to \nu_{\mu})$ disappearance experiments at accelerators and atmospheric experiments. All these measurements are summarized in~\cite{PDG10}.
Here, $\Delta \tilde{m}^2$ is a weighted average of $\Delta m_{31}^2$ and $\Delta m_{32}^2$,  called effective $\Delta m^2$ as described in detail later in this section.  
Recently, finite $\theta_{13}$ was finally measured by short baseline reactor neutrino 
experiments $(\bar{\nu}_e \to \bar{\nu}_e)$~\cite{DC12, DB12, RENO12} and long baseline accelerator  experiments 
$(\nu_{\mu} \to \nu_e)$~\cite{MINOS_th13_11, T2K11}.

Another effective mass squared difference $\Delta \tilde{m}_{31}^2$ can be measured by energy spectrum distortion and baseline dependence of the reactor-$\theta_{13}$ experiments. 
This paper is to measure $\Delta \tilde{m}_{31}^2$ by baseline dependence of the reactor 
neutrino-$\theta_{13}$ experiments.

In reactor-$\theta_{13}$ experiments, usually the neutrino disappearance is analysed by 
a two flavor neutrino oscillation formula; 
\begin{equation}
P(\bar{\nu}_e \to \bar{\nu}_e) 
= 1 - \sin^22\theta_{13}\sin^2 \frac{\Delta\tilde{m}_{31}^2}{4E_{\nu}}L,
 \label{eq:2flavor_oscillation}
\end{equation}
where $L$ is baseline which is $\sim$ 1~km and $E_{\nu}$ is neutrino energy, which 
is around a few MeV. 
$\Delta \tilde{m} _{31}^2$ is a weighted average of the two mass square differences, 
$|\Delta m_{31}^2| $ and $|\Delta m_{32}^2| $ of the standard parametrization, 
\begin{equation}
 \Delta \tilde{m}_{31}^2 = c_{12}^2   \left| \Delta m_{31}^2 \right| 
                          + s_{12}^2 \left| \Delta m_{32}^2 \right|,
\end{equation}
with $c_{ij}$ and $s_{ij}$ representing $\cos\theta_{ij}$ and $\sin\theta_{ij}$, respectively~\cite{Nunokawa05}.
In the analyses of reactor-$\theta_{13}$ experiments so far published, $\sin^22\theta_{13}$ is extracted assuming
$\Delta \tilde{m}_{31}^2 = \Delta \tilde{m}_{32}^2$, which is measured by MINOS experiment~\cite{MINOS_dm32_11}.
$\Delta \tilde{m}_{32}^2$ can be expressed as,
\begin{equation}
 \begin{split}
 \Delta \tilde{m}_{32}^2 
  &= (s_{12}^2 +s_{13}t_{23}\sin2\theta_{12}\cos\delta) \left| \Delta m_{31}^2 \right|
 \\
   &+ (c_{12}^2 - s_{13}t_{23}\sin2\theta_{12}\cos\delta)\left| \Delta m_{32}^2 \right|,
 \label{eq:wmean_Dm2}
 \end{split}
\end{equation}
where $t_{ij} = \tan\theta_{ij}$~\cite{Nunokawa05}.
Since there is a relation
\begin{equation}
 \Delta m_{31}^2 = \Delta m_{32}^2 + \Delta m_{21}^2  ,
 \label{eq:3sum}
\end{equation}
in the standard three neutrino flavor scheme, 
the difference between $\Delta \tilde{m}_{31}^2$ and $\Delta \tilde{m}_{32}^2$ is 
expressed as follows,
\begin{align}
 \frac{2(\Delta \tilde{m}_{31}^2 - \Delta \tilde{m}_{32}^2)}
 {\Delta \tilde{m}_{31}^2 + \Delta \tilde{m}_{32}^2}
 \sim &  \pm (1-s_{13}t_{23}\tan2\theta_{12}\cos \delta)
 \notag \\
 \times \frac{  2\cos2\theta_{12} |\Delta m_{21}^2|}{|\Delta m_{31}^2| + |\Delta m_{32}^2|}
 &\sim \pm 0.012 \times (1 \pm 0.3),
 \label{eq:difference_Dm2}
\end{align}
where the overall sign depends on mass hierarchy, and the $\pm 0.3$ term comes from the ambiguity of $\cos \delta$. The difference is much smaller than the current precisions of measurements and can be treated practically equivalent. A precision better than  1\%  is necessary to distinguish the mass hierarchy. However, if $\Delta\tilde{m}_{31}^2$ and $\Delta\tilde{m}_{32}^2$ are separately measured and if they turn out to be significantly different, it means the standard three flavour neutrino scheme is wrong. Thus it is important to measure $\Delta\tilde{m}_{31}^2$ independently from $\Delta\tilde{m}_{32}^2$ to test the standard three flavour neutrino oscillation.

The $E$ dependence and $L$ dependence analyses to extract $\Delta \tilde{m}_{31}^2$ use independent information, namely energy distortion and normalization and thus are complementary. 
Some of the authors demonstrated $\Delta \tilde{m}_{31}^2$ measurement using $L$ dependence of deficit value of each reactor-$\theta_{13}$ experiment in 2012~\cite{Thiago12,Thiago12_Nu12}.  
In this paper the analysis is significantly improved by applying a detailed global fit making use of the publicly available information of the three reactor neutrino experiments. 

In next section we re-analyze the published data of each experiment 
and compare with the results written in the papers in order to demonstrate our analysis produces identical result. 
Section-III discusses about possible correlations between the experiments. 
 In section-IV, most recent Double Chooz, Daya Bay and RENO results~\cite{DC12, DB12_II, RENO12} are combined and   $\Delta \tilde{m}_{31}^2$ is extracted.
 Finally, a summary of this study is presented in section-V.
 %
  \section{\label{sec:ExpData} Reactor Neutrino Data}
Details of each experiment and their data are presented in this section and they are re-analysed by the authors in order to demonstrate that the analysis methods used in this work are consistent with the publications from the experimental groups. 
The $\chi^2$ used in this section will be used to form global $\chi^2$ function in section-IV. 

  \subsection{Daya Bay}
  \label{sec:DayaBay}
\begin{figure}[htbp]
 \begin{center}
 \includegraphics[height=50mm]{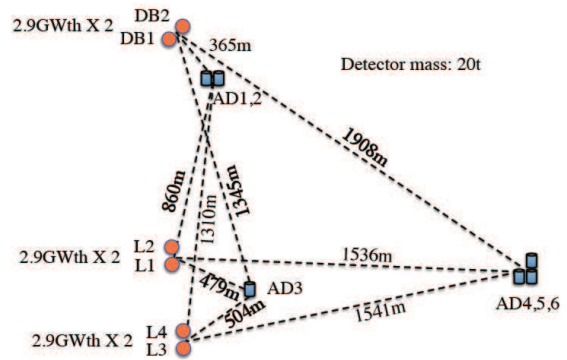}
 \caption{ Relative locations of detectors and reactors of Daya Bay Experiment. 
 Scale is approximate.}
 \label{fig:DB_Location}
 \end{center}
\end{figure}

The Daya Bay (DB) reactor neutrino experiment consists of three experimental halls (EH), containing one or more antineutrino detectors (AD). The AD array sees 6 reactors clustered into 3 pairs: Daya Bay (DB1, DB2), Ling Ao (L1, L2) and Ling Ao-II (L3, L4) power stations. 
Fig.-\ref{fig:DB_Location} shows the relative locations of reactors and AD and table-\ref{tab:DB_Baselines} shows the distance between each combination of reactor and detector.
All reactors are functionally identical pressurized water reactors with maximum thermal power of 
2.9~GW~\cite{DB12}.

In DB publication, the $\chi^2$ is defined as 
\begin{equation}
\begin{split}
 \chi^2_{\rm DB}&(\theta_{13}, \Delta m_{31}^2) = \\  
&\sum_{d}^6 \frac{\left[ M_d  + \eta_d - T_d \left( 1+a + \sum_r^{6}  \omega_r^d \alpha_r + \epsilon_d \right) \right]^2}{M_d + B_d} \\
 & + \sum_r^{6} \frac{\alpha_r^2}{\sigma_r^2}+ \sum_d^6 \left( \frac{\epsilon_d^2}{\sigma_d^2} + \frac{\eta_d^2}{(\sigma_d^b)^2} \right), 
\end{split}
 \label{eq:ChiDB}
\end{equation}
where $M_d$ are the measured neutrino candidate events of the $d$-th AD with background subtracted,
$B_d$ is the corresponding background,
$T_d$ is the prediction from neutrino flux, Monte Carlo simulation (MC) and neutrino oscillation.
$\omega_r^d$ is the fraction of neutrino event contribution of the $r$-th reactor to the $d$-th AD determined by baselines and reactor fluxes.
The uncorrelated reactor uncertainty is $\sigma_r$.
$\sigma_d$ is the uncorrelated detection uncertainty, and $\sigma_d^b$ is the background uncertainty, with the corresponding pull-terms ($\alpha_r, \epsilon_d, \eta_d$). 
An absolute normalization factor $a$ is determined from the fit to the data.

The values of $\omega_r^d$ are not shown in Daya Bay publications and was estimated using 
\begin{equation}
 \omega_r^d = \frac{p_r/L_{rd}^2}{\sum_r(p_r/L_{rd}^2)}
 ~~{\rm with}~~
 p_r =\frac{w_r}{\sum_r w_r},
 \label{eq:DB_Weitedpower}
\end{equation}
where $w_r$ is the thermal power of each reactor and $L_{rd}$ is the baseline of $r$-th reactor 
to $d$-th detector. 
In this analysis, the value of $p_r$ is considered 1/6 since all reactors have same nominal thermal power.
The calculated $\omega_r^d$ is shown in table-\ref{tab:DB_Contributions}. All the others terms are shown in table-\ref{tab:DB_Parameters}.

By using equation-(\ref{eq:ChiDB}) and the data from tables-\ref{tab:DB_Contributions} and -\ref{tab:DB_Parameters}, we were able to reproduce Daya Bay's result, where $T_d$ was multiplied by the value of the deficit probability ($P^{\rm def}_{dr}$), defined as:
\begin{equation}
 P_{dr}^{\rm def} = 1-
 \sin^2 2\theta 
 \frac{\int^{8.0 \rm MeV}_{1.8 \rm MeV}  \sin^2 (1.27 \frac{\Delta m^2 L_{dr}}{E}) n_{\nu}(E)dE}{\int n_{\nu}(E)dE},
  \label{eq:DeficitProbability}
\end{equation}
with  $\Delta m^2$ being measured in eV$^2$, $L_{dr}$ in meters and $E$ in MeV. 
 $n_{\nu}(E)$ is the expected energy spectrum of the observed neutrinos which is calculated by 
 $n_{\nu}(E) = S(E)\sigma_{\rm IBD}(E)$.
 $S(E)$ is the energy spectrum of the reactor neutrinos, which is a sum of the energy spectrum of neutrinos from the four fissile elements:
\begin{equation}
 S(E_{\nu}) = \sum_{i = ^{235}{\rm U},^{238}{\rm U}, ^{239}{\rm Pu}, ^{241}{\rm Pu}}
  \beta_i S_i(E_{\nu}),
 \label{eq:flux}
\end{equation}
where $S_i(E_{\nu})$ is reactor neutrino spectrum per fission from fissile element $i$ and 
$\beta_i$ is a fraction of fission rate of fissile element $i$.
For equilibrium light water reactors, $\beta_i$ are similar and we use the values of Bugey paper~\cite{Bugey94}, namely 
$^{235}{\rm U}: ~^{238}{\rm U}: ~^{239}{\rm Pu}: ~^{241}{\rm Pu}$ 
= 0.538 : 0.078 : 0.328 : 0.056.
 In this study, $S_i(E)$ is approximated as an exponential of a polynomial function which is  defined in~\cite{Mueller11},
\begin{equation}
 S_i(E_{\nu}) \propto \exp \left[ \sum_{j=1}^6 \alpha_j E_{\nu}^{(j-1)}  \right].
 \label{eq:spectrum}
\end{equation}

$\sigma_{\rm IBD}$ is the cross section of the inverse process of neutron $\beta$-decay (IBD), 
that can be precisely calculated from the neutron lifetime~\cite{Vogel99}. 
The energy dependence of the IBD cross section is, 
\begin{equation}
 \sigma_{\rm IBD}(E_{\nu}) \propto (E_{\nu}[{\rm MeV}]-1.29)
 \sqrt{E_{\nu}^2 - 2.59 E_{\nu} +1.4}.
 \label{eq:IBD_xsection}
\end{equation}

\begin{figure}[htbp]
 \includegraphics[height=50mm]{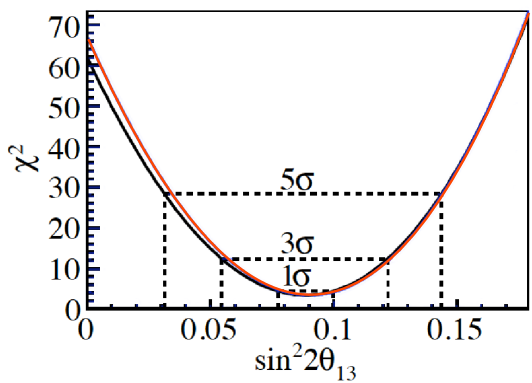}
 \caption{ $\chi^2$ distribution with respect to $\sin^22\theta_{13}$ by fixing $\Delta m^2$ as 
 $\Delta \tilde{m}_{32}^2$ for Daya Bay data. The black curve is the $\chi^2$ distribution shown in their paper \cite{DB12_II} with central value and 1$\sigma$ uncertainty of $0.089 \pm 0.011$, while the red curve shows the $\chi^2$ distribution calculated in this analysis with central value and 1$\sigma$ uncertainty of $0.090^{+0.011}_{-0.010}$.}
 \label{fig:DB_chi2_test}
\end{figure}

$\sin^22\theta_{13}$ is extracted by fixing $\Delta \tilde{m}_{31}^2$ as the MINOS  
$\Delta \tilde{m}_{32}^2 = 2.32 \times 10^{-3}~{\rm eV}^2$~\cite{MINOS_dm32_11}.
The $\chi^2 $ distributions of the Daya Bay paper and our calculation are compared in
fig.-\ref{fig:DB_chi2_test}. 
The Daya Bay central value and uncertainty is $\sin^22\theta_{13\text{DB}} = 0.089 \pm 0.011$ while our analysis showed $\sin^22\theta_{13} = 0.090^{+0.011}_{-0.010}$, in good agreement with the published value. 
We also verified how different values for the fission rates coefficients of equation-(\ref{eq:spectrum}) and different assumptions for equation-(\ref{eq:DB_Weitedpower}), affect the final result. Dependence on the burn-up values is less than 0.001, as it was determined by replacing the burn-up assumption with that of the Chooz reactors at the beginning and end of the reactor cycle. Extreme assumptions on equation-(\ref{eq:DB_Weitedpower}) (one or two reactors off for the whole data period, for example) had an effect of less than 0.002 on the central value, with no change on the sensitivity. Moreover, the good agreement between the $\chi^2$ distributions, shows that the assumptions are reasonable.

%
\subsection{RENO}
\label{sec:RENO}
\begin{figure}[htbp]
 \includegraphics[height=45mm]{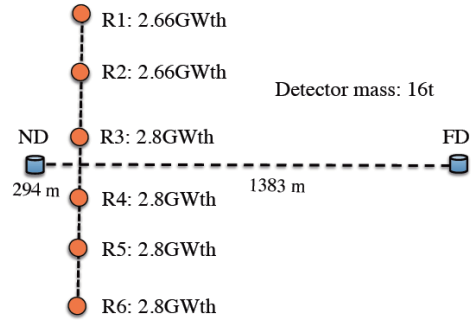}
 \caption{ Relative locations of detectors and reactors of RENO. 
 Scale is approximate.}
 \label{fig:RENO_Location}
\end{figure}

The Reactor Experiment for Neutrino Oscillation (RENO) is located in South Korea and has two identical detectors, one near (ND) and one far (FD) from an 
array of six commercial nuclear reactors, as shown in fig.-\ref{fig:RENO_Location}.

Together with the distances of each detector reactor pair, the contribution of each reactor flux to each detector  for the period of their first analysis is available~\cite{RENO_Nu12} and are summarized in table-\ref{tab:RENO_BL_NF}.

The RENO $\chi^2$ is defined as: 
\begin{equation}
\begin{split}
\chi^2_{\rm RE}&(\theta_{13}, \Delta m_{31}^2) = \\ 
&\sum_{d}^2 \frac{\left[ N_{\rm obs}^d + b_d - (1 + n + \xi_d) \sum_{r}^6 (1+f_r) N_{\rm exp}^{d,r}\right]^2}{N_{\rm obs}^d} 
\\
 &+\sum_{d}^2 \left( \frac{\xi_d^2}{(\sigma_d^{\xi})^2} + \frac{b_d^2}{(\sigma_d^b)^2}\right)
 + \sum_{r}^2 \frac{f_r^2}{\sigma_r^2},
\end{split}
\label{eq:ChiRE}
\end{equation}
where $N_{\rm obs}^d$ is the number of observed IBD candidates in each detector after background subtraction and $N_{\rm exp}^{d,r}$ is the number of expected neutrino events, including detection efficiency, neutrino oscillations and contribution from the $r-$th reactor to each detector determined from baseline distances and reactor fluxes. 
A global normalization $n$ is taken free and determined from the fit to the data.
The uncorrelated reactor uncertainty is $\sigma_r$, $\sigma_d^{\xi}$ is the uncorrelated detection uncertainty, and $\sigma_d^b$ is the background uncertainty, and the corresponding pull parameters are $(f_r,~\xi_d,~b_d)$.
The values of these variables are shown in the table-\ref{tab:RENO_Parameters}.

The expected number of events for both detectors are not present in the RENO paper, but the ratio between data and expectation is shown.
This ratio and the quantities of table-\ref{tab:RENO_BL_NF} were used to calculate the expectation value ($N^{d,r}_{\rm exp}$).

\begin{figure}[htbp]
 \includegraphics[height=35mm]{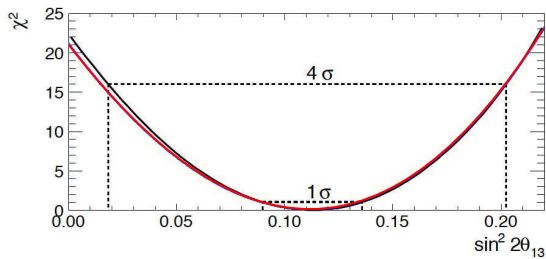}
 \caption{ $\chi^2$ distribution with respect to $\sin^22\theta_{13}$ by fixing $\Delta m^2$ as 
 $\Delta \tilde{m}_{32}^2$ for RENO data.
 The black curve shows the $\chi^2$ distribution shown in their paper \cite{RENO12} with central value and 1~$\sigma$ uncertainty of $0.113 \pm 0.023$, while the red curve shows the 
 $\chi^2$ distribution calculated in this analysis with central value and 1~$\sigma$ uncertainty of $0.111 \pm 0.024$.}
 \label{fig:RENO_chi2_test}
\end{figure}

Using the data in the table-\ref{tab:RENO_Parameters}, equation-(\ref{eq:DeficitProbability}) and the MINOS 
$\Delta \tilde{m}_{32}^2$, we obtained $\sin^22\theta_{13} = 0.111 \pm 0.024$ which is in good agreement with their published value of $\sin^22\theta_{13\text{RE}} = 0.113 \pm 0.023$. 
The $\chi^2$ distributions are also very similar as shown in fig.-\ref{fig:RENO_chi2_test}.
%
\subsection{Double Chooz}
\begin{figure}[htbp]
 \includegraphics[height=25mm]{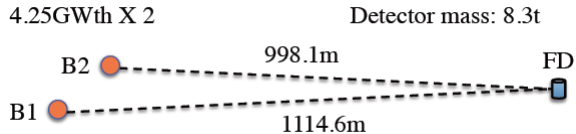}
 \caption{ Relative locations of detector and reactors of Double Chooz experiment.
 Scale is approximate. }
 \label{fig:DC_Location}
\end{figure}

The Double Chooz (DC) experiment uses the two Chooz B reactors with thermal power of 
$4.25~{\rm GW}_{\rm th}$ each. 
Currently, the experiment is using only the far detector, since its near detector is not complete yet.
The Bugey-4 measurement~\cite{Bugey94} is used as a reference of the absolute neutrino flux in the analysis, and the relative location of the far detector and reactors are shown in 
fig.-\ref{fig:DC_Location}, where the distances from the detector to each reactor are 998.1 and 1114.6 meters~\cite{CHOOZ99}.

The Double Chooz collaboration published a rate plus shape analysis result~\cite{DC12}. 
An effect of the shape analysis in this case is an evaluation of main backgrounds of $^9$Li and fast-neutron from the energy spectrum beyond the reactor neutrino energy range.
Since information of detailed energy spectrum, which is necessary to reproduce the analysis, are not publicly
available, we do not consider here the shape analysis but restrict only to the rate analysis.
After the second publication on the $\sin^22\theta_{13}$ measurement, the Double Chooz group
published a result of the direct measurement of backgrounds by making use of 7.53~days reactor-OFF period~\cite{DCoff}.
We used these data in addition to the background evaluation inputs written in~\cite{DC12} to improve the background estimation instead of the energy spectrum analysis.  
The relative neutrino-flux uncertainty for
reactor-OFF period is much larger than reactor-ON period. 
The dominant uncertainty comes from long-life isotopes whose abundance are not well known.
It has negligible contribution in reactor-ON period~\cite{DCoff}. 
Therefore, we regard the error correlation on neutrino flux between the reactor ON and OFF 
periods to be uncorrelated. 
We performed a similar $\chi^2$ analysis as Daya Bay and RENO cases, assuming that the detector and background related uncertainties of \cite{DCoff} and \cite{DC12} are fully correlated. 
\begin{equation}
 \begin{split}
  \chi^2_{\rm DC}&(\theta_{13}, \Delta m_{31}^2) = \\  
  &\sum_{i=1}^2 \frac{\left[ N_i^{\rm obs} - 
  \left(N_i^{\rm exp}( 1+ \alpha_i + \epsilon)+B_i(1+b)\right) \right]^2}
  {N_i^{\rm exp} + B_i} \\
  & + \sum_i^{2} \left(\frac{\alpha_i}{\sigma_r^i} \right)^2+ \frac{\epsilon^2}{\sigma_d^2} +  
  \frac{b^2}{\sigma_b^2}, 
   \label{eq:ChiDC}
 \end{split}
\end{equation}
where $N^{\rm obs}$ is the number of the observed neutrino event candidates. The subscript ``$i$'' represents reactor-ON and OFF period. 
$N^{\rm exp}$ is the number of expected neutrino events, including detection efficiency and oscillation effects, and $B$ is the total expected number of background events. 
The $\sigma_r$, $\sigma_d$, and $\sigma_b$ are the reactor, detection and background uncertainties, respectively. 
The corresponding pull parameters are $(\alpha,~\epsilon,~b)$.
 Using the parameters shown in table-\ref{tab:DC_Parameters}, we obtained 
 $\sin^22\theta_{13} = 0.131 \pm 0.048$ which is consistent with the result of the DC publication, $\sin^22\theta_{13} = 0.109 \pm 0.039$,  although the background evaluation methods are different using different data sets. We also did a rate only analysis of the Double Chooz data, which result agreed with the published one.

\section{\label{sec:Correlation} Correlation Evaluation of Systematic Uncertainties}
In reactor neutrino experiments, the expected number of observed events ($N_{\rm exp}$) is defined by:
\begin{equation}
N_{\rm exp} = \frac{1}{4\pi L^2} N_{\rm p} \varepsilon \frac{P_{\rm th}}{\langle E_f \rangle} \langle \sigma_f \rangle ,
\label{eq:NuInteraction}
\end{equation}
where $L$ is the reactor-detector baseline, $N_{\rm p}$ is the number of targets in the detector, 
$\varepsilon$ is the detector efficiency, $P_{\rm th}$ is the reactor thermal power, $\langle E_f \rangle$ is the mean energy released per fission, and $\langle \sigma_f \rangle$ is the cross-section per fission defined as:
\begin{equation}
\langle \sigma_f \rangle = \sum_i \beta_i \int S_i(E) \sigma_{\rm IBD}(E) dE.
\label{eq:CrossSectionPerFission}
\end{equation}
For each experiment, $L$, $N_{\rm p}$, $\varepsilon$, and $P_{\rm th}$ terms are determined independently. 
Therefore they can be assumed to be uncorrelated. 
On the other hand, $\langle E_f \rangle$ and $\langle \sigma_f \rangle$ terms are taken from the same references and the uncertainties of these terms are correlated between the experiments. 
From the Bugey and Chooz experimental results, the total uncertainty on spectrum prediction is 2.7\%, where a 2\% correlation is expected between the experiments as treated in~\cite{ReactorAnomaly}.
Fully correlated signal prediction uncertainties between experiments, which come from
neutrino flux and detection efficiency, can be cancelled by
overall normalization factors used in the analyses of the Daya Bay and
RENO. 
It allows us only to take into account remaining uncertainties
between detectors or periods for each experiment. Daya Bay and RENO
treat the remaining uncertainties as uncorrelated in their
publications. 
%
\section{\label{seq:GlobalFit} Combined Analysis}
As explained before, the main method of this work is to combine all the data of the current neutrino reactor experiments in a single $\chi^2$ function. 
Then we look for the minimum $\chi^2$ value, calculate the $\Delta \chi^2$ distribution, and determine the confidence level regions.
The $\chi^2$ function used for such analysis was chosen so as to use the data from 
tables-\ref{tab:DB_Baselines} to -\ref{tab:DC_Parameters} 
as well as the correlation as described in section-\ref{sec:Correlation}. 
The definition of our global $\chi^2$ is,
\begin{equation}
\chi^2_{\rm G} \equiv  \chi^2_{\rm DB} + \chi^2_{\rm DC} + \chi^2_{\rm RE},
 \label{eq:GlobalFit}
\end{equation}
with the $\chi^2$ of each experiment defined as in section-II. 
Therefore, this function has 32 pull terms: 18 for Daya Bay (6 reactors, 6 detectors and 6 backgrounds), 10 for RENO (6 reactors, 2 detectors and 2 backgrounds) and 4 for Double Chooz (2 reactors, 1 detector and 1 background). 
It also contains the two overall normalization factors, one for Daya Bay and the other for RENO data set. 

\begin{figure}[htbp]
 \includegraphics[height=60mm]{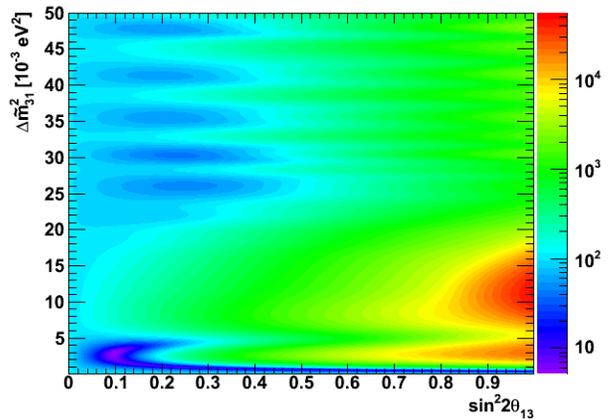}
 \caption{Absolute $\chi^2$ value calculated for each pair of $\Delta \tilde{m}^2_{31}$ and $\sin^22\theta_{13}$, and by the minimization of the pull terms. For higher values of $\Delta \tilde{m}^2_{31}$ (bigger than $10^{-2} \rm eV^2$) some valleys are present, although they are about more than ten times less sensitive than the minimum $\chi^2$.
 }
 \label{fig:absChi2}
\end{figure}

\begin{figure}[htbp]
 \includegraphics[height=60mm]{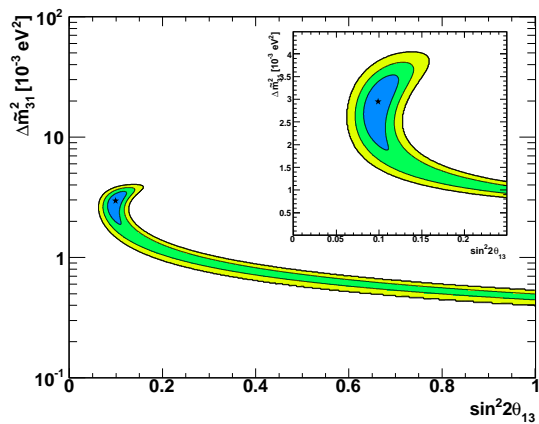}
 \caption{$\Delta\chi^2$ distribution around the $\chi^2_{\rm min}$.
 From the inner to the outer part, the lines correspond to  
 1~$\sigma$, 2~$\sigma$ and 3~$\sigma$ confidence level. 
 The star shows the best fit point.
 There is no solution more significant than 3~$\sigma$ except for the $\chi^2_{\rm min}$ valley.  }
 \label{fig:CombinationContour}
\end{figure}

For all combinations of $\Delta m^2$ and $\sin^22\theta$, the $\chi^2_{\rm G}$ is minimized with respect to the pull terms. 
Fig.-\ref{fig:absChi2} shows a map of the absolute $\chi^2$ and fig.-\ref{fig:CombinationContour} shows the $\Delta \chi^2$ contour map near the $\chi_{\rm min}^2$, obtained by such procedure.
From the minimum point and the 1~$\sigma$ error region in the 1-D $\chi^2$ distribution,
\begin{eqnarray*}
 \chi^2_{\rm min} &=& 5.14\,/\,6 ~{\rm Degrees ~of ~Freedom}, \\
    \Delta \tilde{m}_{31}^2 &=& 2.95^{+0.42}_{-0.61} \times 10^{-3}~{\rm eV}^2, \\
    \sin^22\theta_{13} &=& 0.099^{+0.016}_{-0.012} \notag
\end{eqnarray*}
are obtained. 
This $\Delta \tilde{m}_{31}^2$ is consistent with $\Delta \tilde{m}_{32}^2$ measured by 
accelerator experiments~\cite{MINOS_dm32_11,T2K_dm32_12}, confirming the standard three flavor neutrino oscillation within the error.
The $\sin^22\theta_{13}$ obtained here is independent from $\Delta \tilde{m}_{32}^2$. 
The small $\chi^2_{\rm min}$/DoF means the data from the three reactor neutrino experiments are consistent with each other.


All the pull terms output were within 1~$\sigma$ from the input value, 
and the normalization factors obtained from the fit to the data,
were both less than 1\%.

In fig.-\ref{fig:FIT_result} the baseline dependence of the disappearance probability of each detector is shown, where 
the probability is calculated using the parameters output which give the best fit.
The Double Chooz has a large effect on this $\Delta \tilde{m}_{31}^2$ determination because it locates at a baseline where the 
slope of the oscillation is large. 
In the near future, when the near detector of the Double Chooz experiment starts operation, the accuracy of this $\Delta \tilde{m}_{31}^2$ measurement is expected to improve much. 

\begin{widetext}
\begin{center}
\begin{figure}[htbp]
 \includegraphics[height=45mm]{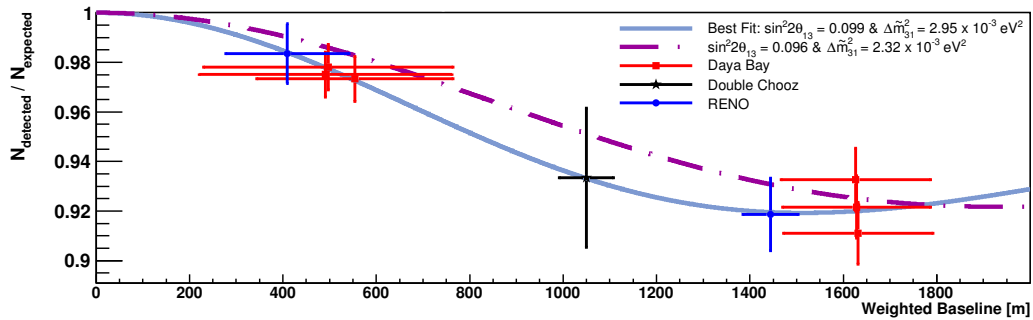}
  \caption{Reactor $\bar{\nu}_e$ survival probabilities.
  The solid line is the oscillation pattern obtained in this analysis and dot-dashed line uses MINOS $\Delta \tilde{m}_{32}^2$ and the $\sin^22\theta_{13}$ that returns the minimum $\chi^2$.
The data points are below the $\Delta \tilde{m}_{32}^2$ because they are calculated using the parameters returned by the best fit solution. 
  Generally, a detector sees several reactors. 
  The horizontal axis is a weighted baseline $\langle L \rangle$ and the horizontal bar in each data point shows the standard deviation of the distribution of the baselines, which is defined by 
  $\sigma_L=\sqrt{\langle L \rangle^2- \langle L^2 \rangle}$, where 
  $\langle L^n  \rangle \equiv \sum_k P_k L_k^{n-2}/\sum_k P_k L_k^{-2}$.
  $k$ is the reactor index and $L_k$ and $P_k$ are the baseline and thermal power of the reactor $k$.
}
 \label{fig:FIT_result}
\end{figure}
\end{center}
\end{widetext}

Complementary to this study, we demonstrated a similar, but simpler and robust measurement of the effective $\Delta \tilde{m}_{31}^2$ from the baseline dependence of the disappearance probabilities of the three reactor-$\theta_{13}$ experiments~\cite{Thiago12, Thiago12_Nu12}. 
The result obtained on that work of 
$\Delta\tilde{m}^2_{31} = 2.99^{+1.13}_{-1.58} \times 10^{-3}~{\rm eV}^2$, 
is compatible with the value obtained in this paper. 
In addition, a similar $\Delta \chi^2$ distribution is presented in~\cite[fig.-4]{ThomasGlobalFit}. However, the central value of $\Delta \tilde{m}_{31}^2$ could not be compared since only the distribution is presented.

\section{Summary}
In this work, a global fit of the data from all the current reactor-$\theta_{13}$ experiments was performed to measure $\Delta\tilde{m}^2_{31}$.
The combination of the data from Daya Bay, RENO and Double Chooz resulted in 
$\Delta\tilde{m}^2_{31} = 2.95^{+0.42}_{-0.61} \times 10^{-3}~{\rm eV}^2$. 
This is consistent with $\Delta\tilde{m}_{32}^2$ and it confirms that the experiments are observing standard three flavor neutrino oscillations within the error. 
The mixing angle obtained this analysis is
$\sin^22\theta_{13} = 0.099^{+0.016}_{-0.012} $. 
The small $\chi^2_{\rm min}/{\rm DoF}$ value indicates that the data from the three reactor experiments are consistent with each other. 
This analysis uses independent information from the energy spectrum distortion and it is possible to improve the accuracy of $\Delta \tilde{m}_{31}^2$ combining with results from energy spectrum analysis. 
It will be important to perform this kind of analysis to improve 
$\Delta \tilde{m}_{31}^2$ accuracy and to check the consistency of the results from the reactor-$\theta_{13}$ experiments.  
%
\section*{\label{sec:Acknowledgement} Acknowledgement}
This work was supported by Ministry of Education, Culture, Sport and Technology of Japan;
Grant-in-Aid for Specially Promoted Research (20001002) and GCOE and focused research project programs of Tohoku University.
We appreciate, Dr.~H.~Nunokawa, Dr.~G.~Mention, Dr.~D.~Lhuillier, the DC and DCJapan groups for useful discussions and comments.
We also appreciate Daya Bay, RENO and Double Chooz members who have taken great effort to produce the precious data.

\clearpage
\begin{widetext}
\begin{center}
\Large{\bf Tables}
\end{center}
\end{widetext}

\begin{table}[htbp]
 \begin{center}
 \caption{   \label{tab:DB_Baselines} Daya Bay: Baselines, in meters, between each detector and core~\cite[tab.-2]{DB12_II}.}
 \begin{ruledtabular}
  \begin{tabular}{crrrrrr}
       & DB1 & DB2   & L1 & L2 & L3 & L4 \\
   \colrule
    AD1         &  362 &  372 &  903 &  817 & 1354 & 1265 \\
   AD2          &  358 &  368 &  903 &  817 & 1354 & 1266 \\
   AD3          & 1332 & 1358 &  468 &  490 &  558 &  499 \\
   AD4          & 1920 & 1894 & 1533 & 1534 & 1551 & 1525 \\
   AD5          & 1918 & 1892 & 1535 & 1535 & 1555 & 1528 \\
   AD6          & 1925 & 1900 & 1539 & 1539 & 1556 & 1530 \\
  \end{tabular}
 \end{ruledtabular}
 \end{center}
\end{table}

\vspace*{8mm}
\begin{table}[htbp]
 \begin{center}
 \caption{ \label{tab:DB_Contributions} Daya Bay: Contribution to each detector from reactor, $\omega_r^d$, calculated using equation-(\ref{eq:DB_Weitedpower}).}
 \begin{ruledtabular}
  \begin{tabular}{crrrrrr}
                & DB1 & DB2   & L1 & L2 & L3 & L4 \\ 
   \colrule
   AD1          &  0.4069 & 0.3854 & 0.0654 & 0.0799 & 0.0291 & 0.0333\\
   AD2          &  0.4089 & 0.3870 & 0.0643 & 0.0785 & 0.0286 & 0.0327\\
   AD3          &  0.0330 & 0.0318 & 0.2676 & 0.2441 & 0.1882 & 0.2354\\
   AD4          &  0.1208 & 0.1241 & 0.1894 & 0.1892 & 0.1851 & 0.1914\\
   AD5          &  0.1201 & 0.1248 & 0.1895 & 0.1895 & 0.1847 & 0.1913\\
   AD6          &  0.1209 & 0.1241 & 0.1892 & 0.1892 & 0.1851 & 0.1914\\
  \end{tabular}
 \end{ruledtabular}
 \end{center}
\end{table}

\begin{widetext}
\begin{center}
\begin{table}[htbp]
 \caption{  \label{tab:DB_Parameters} Daya Bay: Fitting parameters. Differently from~\cite{DB12_II}, here the efficiency and backgrounds (BKG) are combined in a single quantity for each detector. The total BKG is subtracted from the IBD candidates giving $M_d$.}
 \begin{ruledtabular}
  \begin{tabular}{crrrrrr}
       & AD1 & AD2   & AD3 & AD4 & AD5 & AD6 \\ 
   \colrule
      $\nu$ candidate & 69121   & 69714 & 66473 & 9788   & 9669   & 9452   \\
    $T_{d0}$          & 68613   & 69595 & 66402 & 9922.9 & 9940.2 & 9837.7 \\
Total BKG [day$^{-1}$]& $13.68 \pm 1.54$ & $13.55 \pm 1.54$ & $10.38 \pm 1.17$ 
                    & $3.56 \pm 0.24$ & $3.55 \pm 0.24$ & $3.44 \pm 0.24$ \\
   Live Time [days]   & 127.5470 & 127.5470 & 127.3763 & 126.2646 & 126.2646 & 126.2646 \\
   Efficiency         & 0.8015 & 0.7986 & 0.8364 & 0.9555 & 0.9552 & 0.9547 \\
   $M_d$              & 67723.59 & 68334.17 & 65363.96 & 9358.7 & 9240.98 & 9037.24 \\
   $\sigma_b$         & 157.43 & 156.86 & 124.65 & 28.95 & 28.94 & 28.93 \\
   $\sigma_d$         & 0.002 &  0.002 & 0.002 & 0.002 & 0.002 & 0.002 \\
   $\sigma_r$         & 0.008 &  0.008 & 0.008 & 0.008 & 0.008 & 0.008 \\
  \end{tabular}
 \end{ruledtabular}
\end{table}
\end{center}
\end{widetext}

\begin{table}[htbp]
 \caption{ \label{tab:RENO_BL_NF}
 RENO: Baselines and neutrino flux contributions~\cite[page-7]{RENO_Nu12}.
 }
\begin{ruledtabular}
  \begin{tabular}{lrrrrrr}
      & R1     & R2     & R3     & R4     & R5     & R6 \\
   \colrule
   FD baseline [m] & 1556.5 & 1456.2 & 1395.9 & 1381.3 & 1413.8 & 1490.1 \\
   FD contribution & 0.1373 & 0.1574 & 0.1809 & 0.1856 & 0.1780 & 0.1608 \\
   \hline
   ND baseline [m]&  667.9 &  451.8 &  304.8 &  336.1 &  513.9 &  739.1 \\
   ND contribution & 0.0678 & 0.1493 & 0.3419 & 0.2701 & 0.1150 & 0.0558 
  \end{tabular}
 \end{ruledtabular}
\end{table}

\begin{table}[htbp]
 \caption{  \label{tab:RENO_Parameters}
 RENO: Fitting parameters. Differently from~\cite{RENO12}, here the BKGs are summed into a single quantity. The total BKG is subtracted from the IBD candidates giving $N_{\rm obs}$. $N^d_{\rm exp}$ is calculated as described in section~\ref{sec:RENO}.}
\begin{ruledtabular}
  \begin{tabular}{lrr}
      & ND     & FD  \\
   \colrule
   IBD candidates   & 154088           & 17102            \\
   $N^d_{\rm exp}$  & 151723.54        & 17565.72         \\
Total BKG [day$^{-1}$]  & $21.75 \pm 5.93$ & $4.24 \pm 0.75$  \\
   Live Time [days] & 192.42           & 222.06           \\
   Efficiency       & 0.647            & 0.745            \\
   $N_{\rm obs}$    & 149902.86        & 16160.46         \\
   $\sigma_d$       & 0.002            & 0.002            \\
   $\sigma_r$       & 0.009            & 0.009            \\
   $\sigma_b$		& 1141.05		   & 166.54
  \end{tabular}
 \end{ruledtabular}
\end{table}

\begin{table}[htbp]
 \caption{  \label{tab:DC_Parameters} Double Chooz: Fitting parameters~\cite{DC12,DCoff}. The detector uncertainty is the combination of detector response and efficiency uncertainties, and the BKGs are combined in a single quantity for each data set.}
\begin{ruledtabular}
  \begin{tabular}{lrr}
      & Reactor-On      & Reactor-Off\\
   \colrule
   IBD candidates   & 8249             & 8 \\
   IBD prediction   & 8439.6           & 1.42\\
   Total BKG [day$^{-1}$]  & $2.18 \pm 0.58$  &  $2.00 \pm 0.58$\\
   Live Time [days] & 227.93           & 6.84 \\
   $\sigma_d$       & 0.010          & 0.010\\
   $\sigma_r$        & 0.017          & 0.40
  \end{tabular}
 \end{ruledtabular}
\end{table}

\clearpage


\end{document}